\documentclass[12pt]{article}
\usepackage{latexsym}
\usepackage{amssymb}
\usepackage{amsmath}
\usepackage{epsf}

\allowdisplaybreaks

\newcommand{\del}[1]{\partial_{#1}}
\newcommand{\non}{\nonumber\\}
\newcommand{\PB}[2]{\left\{ #1 , #2 \right\}}
\newcommand{\DB}[2]{\left\{ #1 , #2 \right\}_{\rm DB}}

\newcommand{\HT}{H_{\rm T}} 
\newcommand{\we}{\approx} 
\newcommand{\combi}[2]{\ensuremath{ \begin{pmatrix} #1 \cr #2 \cr
          \end{pmatrix}}} 
\newcommand{\trs}[1]{#1 ^{\rm T}}
\newcommand{\deldash}{\delta^{\prime}}
\newcommand{\Xd}[1]{X_0^{#1 \prime}}
\newcommand{\Pd}[1]{P_0^{#1 \prime}}
\newcommand{\Xp}[1]{X^{\prime #1}}

\newcommand{\ph}[1]{\phantom{#1}}

\begin{document}

\vspace{10mm}
\begin{titlepage}
\title{
\hfill\parbox{4cm}
{\normalsize KUNS-1664\\{\tt hep-th/0005123}}\\
\vspace{1cm}
Open membranes in a constant $C$-field background \\
and noncommutative boundary strings
}
\author{
Shoichi {\sc Kawamoto}\thanks{{\tt kawamoto@gauge.scphys.kyoto-u.ac.jp}}
{} and 
Naoki {\sc Sasakura}\thanks{{\tt sasakura@gauge.scphys.kyoto-u.ac.jp}}
\\[7pt]
{\it Department of Physics, Kyoto University, Kyoto 606-8502, Japan}
}
\date{\normalsize May, 2000}
\maketitle
\thispagestyle{empty}

\vspace{10mm}

\begin{abstract}
\normalsize\noindent 
We investigate the dynamics of open membrane boundaries in a constant
$C$-field background. 
We follow the analysis for open strings in a $B$-field background,
and take some approximations. 
We find that open membrane boundaries do show
noncommutativity in this case by explicit calculations.
Membrane  boundaries are one dimensional strings,
so we face  a new type of noncommutativity, that is, noncommutative strings.
\end{abstract}

\end{titlepage}


\section{Introduction}
\label{sec:introduction}

It is surprising that,
although it seems that noncommutative geometry is quite a pure
mathematical object, noncommutativity does emerge in
some definite limits of string theory. 
For instance, matrix theory compactified on tori gives Yang-Mills
theory on noncommutative tori\cite{CDS97}; 
the quantization of open strings on a
D-brane with a background $B$-field leads this D-brane world-volume to
become noncommutative\cite{CH98}; 
the twisted version of the reduced large-$N$ Super Yang-Mills
model originally  
considered as a constructive definition of type IIB superstring
can be interpreted as noncommutative Yang-Mills
theory\cite{IIBnoncomm}, and so on.

Recent development on string dualities reveals that M-theory rules
nonperturbative features of superstring theories.
It is natural to ask what is noncommutativity in M-theory.
We do not know so much about M-theory. 
M-theory leads to eleven dimensional supergravity at the low-energy limit,
and  M-theory compactified on a circle 
becomes type IIA superstring by taking the limit for the radius of the circle
to become zero.
Moreover M-theory  contains the two-dimensional extended object, M2-brane,
 as the fundamental component.
Matrix theory proposed by Banks, Fischler, Shenker and Susskind
\cite{BFSS} is considered as describing some (or complete
as they state originally) degrees of freedom of M-theory.
This matrix theory does show noncommutativity in some cases
commented above.
We can expect naturally  that noncommutativity can emerge in M-theory.

On the other hand, 
a supersymmetric two-dimensional extended object, called
supermembrane, is interesting in its connection to superstrings.
A quantum extension of supermembrane is expected to give a definition
of M-theory.
Especially, it is well known that supermembrane in eleven dimensions
can consistently couple to eleven dimensional supergravity as its
backgrounds\cite{BST}.
Thus, we have a natural question here; how does supermembrane theory show
noncommutativity?
It is a very meaningful question in two reasons. First, since we expect that
supermembrane is a definition of M-theory, we also expect that
supermembrane theory has noncommutativity in a definite limit or a
background. 
Secondly, we wonder what is noncommutativity in  more than
two-dimensional extended objects. 
To clear this second point, let us compare it with the string case.
In string theory, the end of open strings becomes noncommutative and
a D-brane world-volume on which open strings can end has
noncommutative geometry.
Then, let us consider an \textit{open membrane} which has
one-dimensional boundary and focus on the behavior of these boundaries.
Here, we face a conceptual jump.
In string theory, open string ends are ``points'' and on a D-brane
world-volume points do not commute with each other, while in membrane case, 
we find that its boundaries are ``strings'' and noncommutativity means
one-dimensional strings do not commute with each other.
Thus, we can learn a new feature of noncommutative geometry by studying
membrane noncommutativity. A primitive analysis was carried out in 
\cite{CH98}.

In string theory, we can find noncommutativity by quantizing open strings
in background NS-NS fields.
Some authors have applied the Dirac procedure to boundary
conditions\cite{CH99,AAS-J99}. 
This method is very transparent and can be easily extended to other systems.
We attempt to investigate an open membrane in a background
three-form field in this way.
 It is well known that to
investigate membrane theory has severe difficulties, for example,
non-linearity of world-volume theory, non-renormalizability of
three-dimensional sigma model, and so on.
Thus, we must take an appropriate approximation, as explained later.

Our plan of investigation is as follows. 
In seeing the noncommutativity,
supersymmetry was not essential in the string case. We drop the fermionic
parts and consider a bosonic membrane.
We start with a bosonic open
membrane in a constant gauge field background.
Since we should take our bosonic membrane as a toy model of
eleven dimensional supermembrane, we restrict the background fields to 
the massless bosonic fields of eleven dimensional supergravity,
the metric $g_{\mu\nu}$ and the three-form tensor field $C_{\mu\nu\rho}$.
We consider only a bosonic background and drop the fermionic field,
the gravitino $\chi^\mu$.
Without introducing a two-form gauge field, there can not 
exist open membranes by gauge-invariance.
Also in supermembrane case, we can not introduce an open supermembrane without 
braking all the supersymmetries in \textit{flat} Minkowski space-time.
However we can formulate a supersymmetric open supermembrane when there
exists a ``topological defect'' as a background \cite{openMem}.
These defects are interpreted as, for instance, M5-brane, ``end of the
world'' 9-plane in Ho\v rava-Witten's sense,  etc.
 We shall introduce fixed $p$-branes in this bosonic case.
We assume our open membranes are bounded to these
``boundary planes,'' and there is a two-form field, to
which open membrane boundaries can couple, on these planes\footnote{%
In \cite{CS97}, an open membrane probe was used to derive the
equations of motion of boundary $M5$-branes.}.
In these settings, we calculate the Dirac brackets and 
confirm noncommutativity on these boundary planes.
Our calculation is only to second order in $C$ and not exact.

This paper is organized as follows.
In section \ref{sec:setup}, we propose our setup.
We consider a bosonic open membrane in a constant $C$-field background.
We suppose that one direction of the target space is compactified
to a circle, another direction is compactified to 
an interval and there exist two fixed planes at the boundaries of this
direction.
We fix the reparametrization invariance of the world-volume 
with a static gauge and simplify the action by taking a limit.
Equations of motion and boundary conditions are found, we go on to
the canonical formalism and impose the boundary conditions as
constraints.
In section \ref{sec:solve},
we solve the constraints with an approximation.
We take the radius of the compactification circle to be very large
and the distance between the boundary planes to be infinitesimally small. 
In section \ref{sec:comp},
we calculate the Dirac brackets and confirm the
noncommutativity on the boundary planes.
Section \ref{sec:conc} is served to discussions and remarks.
In appendix \ref{app:Dirac},
 we review the application of Dirac's procedure for constrained
 systems to the boundary constraints in  the string case.

\section{An open membrane in a constant $C$-field}
\label{sec:setup}

Let us consider an open membrane in the background of a constant three-form 
tensor field $C_{\mu\nu\rho}$.
We suppose that our membrane topology is cylindrical and the
background is eleven dimensional,
compactified to 
$\textbf{R}^{9-p} \times M^p \times S^1 \times I$,
 where $M^p$ is a $p$
dimensional flat Minkowski space-time and $I$ is
an interval with a finite length\footnote{%
Conventions of indices are as follows.
$\mu,\nu,\cdots$ are eleven dimensional
suffices and $i,j,\cdots$ represent the spatial directions of 
the $p$-brane world-volume. 
Membrane world-volume indices are $\alpha , \beta ,\ldots$ and
$a,b$ are world-volume spatial indices, $a,b = 1,2$.}.
There exist at the boundaries of $I$ two $p$-branes on which
an open membrane can end, and the $p$-branes wrap once around the $S^1$.
$\textbf{R}^{9-p} \times I$ is transverse to these $p$-branes.   
We drop the fermionic part, that is, 
restrict ourselves to considering a bosonic
membrane. 

\begin{figure}[hbt]
\begin{center}
\leavevmode
\epsfxsize=120mm
\epsfbox{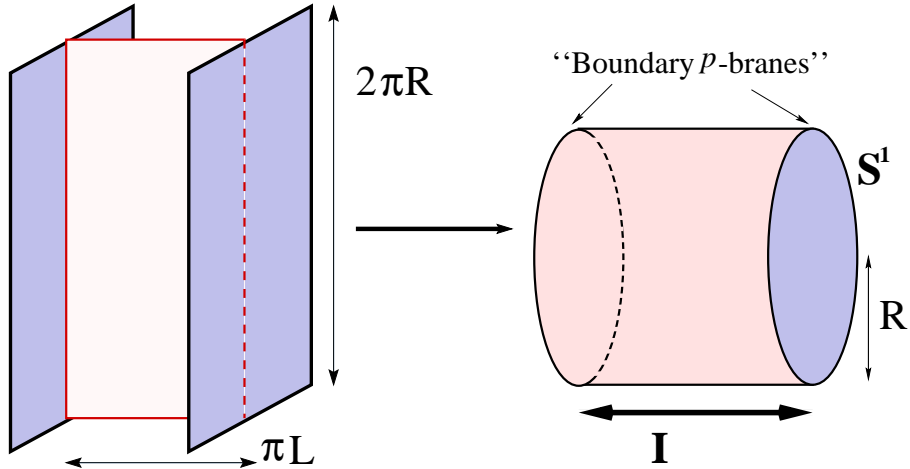}
    \caption{A membrane wrapped once around the compactification circle stretches  between two fixed $p$-branes.}
\label{fig:setup}
\end{center}
\end{figure}

In this case, the action of the membrane is
\begin{equation}
   \label{open+Cfield1}
  S = -T \int \!\! d^3\xi \left\{ \sqrt{- \det h_{\alpha \beta}} 
  + \frac{1}{3!} \epsilon^{\alpha\beta\gamma} C_{\mu\nu\rho}
  \del{\alpha } X^\mu \del{\beta } X^\nu  \del{\gamma} X^\rho \right\}
 , 
\end{equation}
where $\xi^\alpha$ are the world-volume coordinates $(\tau,\sigma_1,\sigma_2)$ and $h_{\alpha\beta}$ is
the induced metric on the world-volume, 
$h_{\alpha\beta} \equiv \del{\alpha}X^\mu \del{\beta} X_\mu$.

First, we fix the gauge freedom of world-volume
reparametrization invariance with the static gauge,
\begin{equation}
\left\{ 
  \begin{array}{lcl}
  X^0  =  \tau  & &  \tau \in (-\infty , \infty
  ) \\
   X^9 = \sigma_1 L  & &  \sigma_1 \in [0 , \pi
  ] \\
  X^{10} = \sigma_2 R  & &  \sigma_2 \in [0 ,
  2\pi ) \ , \\
\end{array} \right.
\end{equation}
and the radius of the compactified direction $X^{10}$ is $R$,
\begin{equation}
X^{10} \sim X^{10} + 2\pi R .
\end{equation}

We also compactify the $X^9$ direction on an interval.
Suppose that there are two  ``fixed planes'' placed at
a distance of $\pi L$ in the $X^9$ direction.
Here, $\pi L$ is the length of the interval, and
the two boundaries of a membrane are bound to each of these ``fixed planes'',
\begin{equation}
  \Delta X^9 = \pi L .
\end{equation}
These ``fixed planes'' are, for example, regarded as M5-branes in
M-theory when $p=5$. 
Since the dimension of the $p$-brane is not essential in our analysis, 
we assume $p=9$ from now on.

Under the static gauge condition, 
\begin{eqnarray}
  \label{deth}
  \det h & = & \left|
    \begin{array}{ccc}
-1 + ( \dot{X} ^i)^2 & \dot{X}^i \del{1} X^i &  \dot{X}^i \del{2} X^i
\\
\dot{X}^i \del{1} X^i & L^2 + (\del{1}X^i)^2 & \del{1}X^i \del{2}X^i
\\
\dot{X}^i \del{2} X^i & \del{1}X^i \del{2}X^i & R^2 + (\del{2} X^i)^2 \\
    \end{array}
\right| \non
& = & -L^2 R^2 + L^2 R^2 (\dot{X}^i)^2 - R^2 (\del{1}X^i)^2 -
L^2(\del{2}X^i)^2 + {\cal O}\left( (\partial X)^4 \right) ,
\end{eqnarray}
and we get the first part of the action (Dirac part) as
\begin{equation}
  S_{\rm D} = T \int \!\! d^3\xi \left[ -1 + \frac{1}{2}(\dot{X}^i)^2
  -  \frac{1}{2}(\del{1}X^i)^2 - \frac{1}{2}(\del{2}X^i)^2 + {\cal
  O}\left( (\partial X)^4 \right) \right] ,
\end{equation}
where we have made a rescaling, 
$L\sigma_1 \rightarrow \sigma_1 \ , \ R\sigma_2
\rightarrow \sigma_2$.

Next, we go on to consider the $C$-field part,
\begin{equation}
  S_C = \int_{\Sigma} \!\! C_{[3]},
\end{equation}
where $\Sigma$ is the world-volume of a membrane.
At the beginning, note
that our action (\ref{open+Cfield1}) is {\it not\/}
gauge-invariant for an {\it open\/} membrane.
So as to make an open membrane
gauge-invariant, we introduce a two-form gauge field $B$
coupled to the boundaries of a membrane,
\begin{equation}
  S_B = \int_{\partial \Sigma} B_{[2]} ,
\end{equation}
which transforms as $B \rightarrow B - \Lambda$ under 
the $C$-field gauge transformation, $C \rightarrow C + d\Lambda$,
where $\Lambda$ is a two-form field.
Here, this $B$-field is on the boundary planes and has the field strength 
$F \equiv dB$ on these planes.
Gauge-invariance requires that $C$ and $F$ always 
appear with the form of $C+F$, so
the constant $C$-field leads to a constant field strength $F$ on 
the boundary planes.
Then, we gauge away $F$ and only consider the effects of the $C$-field.
Moreover, we suppose
that the $C$-field is not only constant but also ``magnetic'', that is, 
their non-zero components are only $C_{ijk}$. 
Finally, the $C$-field part of the action is
\begin{equation}
\label{C-part}
  S_{C} = -T \int \!\! d^3 \xi \, 
C_{ijk} \dot{X}^i \del{1} X^j \del{2} X^k ,
\end{equation}
where we have made a rescaling $C \rightarrow (LR)^{-1} C$.

A part of difficulties of membrane theory comes from its non-linearity 
of world-volume theory.
Here, to avoid it, we take the limit $\alpha \rightarrow \infty$,
\begin{align}
T &\rightarrow \alpha^2 T, \non
X &\rightarrow \frac{1}{\alpha} X, \non
C &\rightarrow \alpha C, \nonumber
\end{align}
and also drop the constant term of the Dirac part.
This limit  means that 
the self-interactions of the world-volume theory are weak
compared to the interactions with the background gauge fields.
Finally, we get the effective action as follows,
\begin{equation}
 \label{effective action}
 S^{\rm  eff} = T \int \!\! d^3 \xi \left[ \frac{1}{2} 
  \left\{ (\dot{X}^i)^2 - (\del{1}X^i)^2 - (\del{2}X^i)^2 \right\} - C_{ijk}
  \dot{X}^i \del{1} X^j  \del{2}  X^k \right] \ ,
\end{equation}
where the ranges of the world-volume coordinates are
\begin{align}
\sigma_1 &\in [0, \pi L] , \\
\sigma_2 &\in [0, 2\pi R ) ,
\end{align}
and the area of the membrane is $2 \pi^2 LR$.

To find the equations of motion and the boundary conditions, we vary the
effective action (\ref{effective action}),
\begin{align}
 \delta S^{\rm  eff} &= -T \int \!\! d^3 \xi 
  \left[ \ddot{X}^i - (\del{1})^2X^i - (\del{2})^2X^i \right] \delta X^i
 \non
& \phantom{aa} + T \int \!\! d^3 \xi \del{1} \left[ \left( -\del{1}X^i
  - C_{ijk}  \dot{X}^k \del{2} X^j \right) \delta X^i \right] \ .
\end{align}
$\delta S^{\rm  eff} =0$ leads to the equations of motion,
\begin{equation}
  \label{EOM}
  \square X^i =0  ,
\end{equation}
where $\square \equiv \eta^{\alpha\beta}
\partial_{\alpha}\partial_{\beta} = \del{\tau}^2 - \del{1}^2 -
\del{2}^2$, 
and also leads to the boundary conditions,
\begin{equation}
  \label{boundary condition}
\left. \del{1}X^i - C_{ijk}\dot{X}^j\del{2}X^k
\right|_{\sigma_1=0,\pi L} = 0 .
\end{equation}

The conjugate momenta are 
\begin{equation}
  \label{conjugate mom}
  P_i = \frac{\delta}{\delta \dot{X}^i}L = T \left( \dot{X}_i -
  C_{ijk} \del{1}X^j\del{2}X^k \right) \ , 
\end{equation}
so the Hamiltonian is 
\begin{align}
  H & \equiv  
\int\!\! d^2\sigma \left( \dot{X}^iP_i - {\cal L} \right) \non
  & =  \frac{T}{2} \int\!\! d^2 \sigma  \left[ \left( \frac{P^i}{T} 
    + C_{ijk} \partial _1 X^j \partial _2 X^k \right )^2 +
    ( \partial _1 X^i )^2 +  ( \partial _2 X^i )^2 \right].
\end{align}

To follow the calculations in the string case\cite{CH99,AAS-J99},
 we regard the boundary conditions as primary constraints,
\begin{equation}
\label{primaryconstraint}
\phi^i _1 =  \left. \partial_1 X^i - C_{ijk} \left( \frac{P^j}{T} +
    C_{jlm}\del{1}X^l\del{2}X^m  \right) \partial _2 X^k \right|
_{\sigma_1 = 0 , \pi L} \approx  0 \ .
\end{equation}

Poisson brackets are ordinarily defined as 
\begin{equation}
  \label{Poisson bracket}
  \begin{array}{l}
  \PB{X^i(\sigma_1,\sigma_2)}{P_j(\sigma'_1,\sigma'_2)} = \delta^i_j
  \delta^2(\sigma-\sigma') , \\
\PB{X^i}{X^j} = \PB{P_i}{P_j} = 0 .
  \end{array}
\end{equation}
Using these, we get the equations of motion,
\begin{equation}
\label{HeomofX}
\dot{X}^i \equiv \PB{X^i(\sigma)}{H} =  \frac{P^i}{T} + C_{ijk} \partial _1 X^j \partial _2  X^k ,
\end{equation}
and
\begin{align}
\dot{P}^i \equiv  \PB{P_i(\sigma)}{H} &= T \left\{ \ddot{X}^i -
 C_{ijk} \left( \del{1} \dot{X}^j\del{2}X^k + \del{1}X^j \del{2}
 \dot{X}^k \right) \right\} \non 
& =  T  \left[  C_{ijk} \left
    ( \partial _2 X^j  \partial_1 \left( \frac{P^k}{T} + C_{klm}
      \partial _1 X^l \partial  _2 X^m \right) \right.  \right. \nonumber\\
 & \phantom{aaaa}  \left. \left.
 -  \partial _1 X^j  \partial_2 \left( \frac{P^k}{T} +
  C_{klm} \partial _1 X^l \partial _2 X^m \right) \right)
   + \Delta X^i  \right] ,
\end{align}
where Laplacian $\Delta$ is defined as $\partial_1^2 +
\partial_2^2$ and dot means $\tau$ derivative.

For simplicity, we set $T=1$. 
We can recover $T$ by replacing  $P$ with $P/T$.

\section{Solving constraints}
\label{sec:solve}

The method described in appendix \ref{app:Dirac} 
leads us to find the Dirac
brackets of the membrane in the constant $C$-field. 
First, we consider the consistency conditions of the constraints
\begin{equation}
  \dot\phi \equiv \PB{\phi}{\HT} \we 0 \ ,
\end{equation}
and find an infinite chain of secondary constraints as follows
\begin{align}
  \phi_2^i & \equiv  \dot{\phi}_1^i = \left\{ \phi_1^i , H \right\}
  \non
  & =  \del{1} \dot{X}^i - C_{ijk} \ddot{X}^j \del{2}X^k - C_{ijk}
  \dot{X}^j \del{2} \dot{X}^k \ , \non
 \phi_3^i & \equiv \dot{\phi}_2^i \non
 & =  \del{1} \ddot{X}^i - C_{ijk} \left[ X^{{ (3)}j} \del{2} X^k
  + 2\ddot{X}^j \del{2} \dot{X}^k + \dot{X}^j \del{2} \ddot{X}^k
  \right] \ , \non
 & \vdots  \non
 \phi_{n+1}^i & \equiv \phi^{{ (n)} i}_1 \non
 & =  \del{1} X^{{ (n)} i}_1 - C_{ijk} \sum^{\infty}_{\ell=0}
  \combi{n}{\ell} X^{{ (n+1-\ell)\/}j} \del{2} X^{{ (\ell)\/}k}
  \ ,
\end{align}
where 
\begin{equation}
\phi^{{ (n)} i} \equiv  \frac{\partial^n}{\partial
  \tau^n} \phi^i  \ .
\end{equation}
Note that the equation of motion (\ref{HeomofX}) tells that each secondary
constraint has  at most $C^3$, and all the  constraints are
 second class.
Explicit computations show that the first few constraints are given by
\begin{align}
\label{eq:constraint1}
\phi^i _1 &=  \left. \partial_1 X^i - C_{ijk} \left( P^j +
    C_{jlm}\del{1}X^j\del{2}X^k  \right) \partial _2 X^k \right|
_{\sigma_1 = 0 , \pi L} \approx  0 , \\
\label{eq:constraint2}
\phi^i _2 &=  \del{1} P^i \nonumber\\
        &   + C_{ijk} \left[ \del{1} X^j \del{1} \del{2}
  X^k - \del{2}^2 X^j \del{2} X^k - P^j \del{2} P^k \right]
  \nonumber\\
         &    + C_{ijk}  C_{jlm} \left[ -\del{2} P^k \del{1} X^l
  \del{2} X^m + P^k \del{2}  (\del{1} X^l \del{2} X^m) \right]
  \nonumber\\
          &  \left. - C_{ijk}C_{jlm}C_{kop} [ \del{1}X^l   \del{2} X^m \del{2}
  ( \del{1} X^o \del{2} X^p ) ] \right|
_{\sigma_1 = 0 , \pi L} \approx  0 , \\
  \phi^i_3 & =  \del{1} \Delta X^k \nonumber\\
     &  + C_{ijk} [ -\Delta P^j \del{2} X^k + 2 \del{2} P^j
     \Delta X^k - P^j \del{2} \Delta X^k ] \nonumber\\
     &  + C_{ijk}C_{jlm} \left[ 
     2 \Delta X^k \del{2} (\del{1} X^l \del{2} X^m ) \right. \non
&  \left. \phantom{+ C_{ijk}C_{jlm} [ 2 \Delta]} \left.
- \del{2} X^k \Delta ( \del{1} X^l \del{2} X^m ) - \del{2}
     \Delta X^k (\del{1} X^l \del{2} X^m) \right] \right|
_{\sigma_1 = 0 , \pi L} \approx  0 .
\end{align}
These constraints look too hard to solve completely unlike 
the string case. Thus, we shall take an approximation to solve them.

At this stage, we take the limit $L \rightarrow 0$ and 
$R \rightarrow \infty$
\footnote{Note that this limit is a tensionless string limit in
  Strominger's sense \cite{Strominger:1996ac}.}. 
This leads to simplification as follows.
For $\sigma_1$, we suppose that no oscillations are excited.
Hence, after solving the constraints, $X^i(\tau, \sigma_1 , \sigma_2)$
 and $P^i(\tau, \sigma_1 , \sigma_2)$ are determined by their boundary 
 values.
And for $\sigma_2$, we neglect terms which is of order $(1/R)^3$ or
higher, which means that we drop the terms involving three
derivatives of $\sigma_2$ or higher,
\begin{equation}
\label{approx:del2}
 \del{2}^3 X^i = 0 \ , \ \del{2}^2 X^i \del{2} X^j =0 \quad \text{etc$\ldots$}
 \ .
\end{equation}

To solve the constraints,
we shall include the effects of the $C$-field order by order.
At order of $C^0$, the boundary conditions are
\begin{equation}
\left. \del{1} X^i \right|_{\sigma_1 = 0,\pi L} = 0 \quad \text{and}
 \quad
X^i(\tau,\sigma_1,\sigma_2) =X^i(\tau,\sigma_1 ,\sigma_2 + 2\pi R) .
\end{equation}
Since no oscillations of $\sigma_1$ are excited under the 
$L \rightarrow 0$ limit, the solution is 
\begin{equation}
  \label{mode exp of X}
  X^i(\tau,\sigma_1,\sigma_2) =  x^{(0)i}_0(\tau,\sigma_2) .
\end{equation}
where the subscript $0$ of $x^{(0)i}_0$ 
means we are considering only the zero-mode of
$\sigma_1$.
Since the $C$-field background changes the $\sigma_1$ boundary
conditions,
the $\sigma_1$ dependence of fields $X$ and $P$ would be
altered:
\begin{equation}
  X^i = x^{(0)i}_0(\tau,\sigma_2) 
+ (\text{corrections which depend also on $\sigma_1$
  and $C$}) .
\end{equation}

Let us calculate the corrections to  second order in $C$.
Consider the expansions of $X$ and $P$ in terms of  $C$
\begin{align}
X^i_0 (\tau,\sigma_1,\sigma_2) = x_0^{(0)i} + x_0^{(1)i} + x_0^{(2)i}
, \\
P^i_0 (\tau,\sigma_1,\sigma_2) = p_0^{(0)i} + p_0^{(1)i} + p_0^{(2)i} ,
\end{align}
where $x_0^{(0)}$ and $p_0^{(0)}$ are functions of $\tau$ and
$\sigma_2$, independent of $\sigma_1$ and unconstrained.
We  substitute them into the constraints
(\ref{eq:constraint1}) and (\ref{eq:constraint2}).
Of order $C^1$, we get
\begin{align}
  \phi_1^i &= \left. \del{1}x_0^{(1)i} - C_{ijk}p_0^{(0)j} \del{2}x_0^{(0)k}
  \right|_{\sigma_1 = 0,\pi L} \approx 0 , \non
  \phi_2^i &= \left. \del{1}p_0^{(1)i} + C_{ijk} 
\left( -p_0^{(0)j} \del{2} p_0^{(0)k}  \right)
  \right|_{\sigma_1 = 0,\pi L} \approx 0 ,
\end{align}
and find solutions at this order as follows
\begin{align}
x_0^{(1)i}(\tau,\sigma_1,\sigma_2)  =
A_0^{(1)i}(\tau,\sigma_2) + C_{ijk} p_0^{(0)j} \del{2}x_0^{(0)k} \cdot
\sigma_1, \\
p_0^{(1)i}(\tau,\sigma_1,\sigma_2) =
B_0^{(1)i}(\tau,\sigma_2) + C_{ijk} p_0^{(0)j} \del{2} p_0^{(0)k}
\cdot \sigma_1 ,
\end{align}
where $A_0$ and $B_0$ in the right hand sides are unconstrained.
In succession, the equations of order $C^2$ are
\begin{align}
  \phi_1^i 
& = \del{1}x_0^{(2)i} - C_{ijk} \left[ p_0^{(1)j}\del{2}x_0^{(0)k} +
  p_0^{(0)j}\del{2}x_0^{(1)k} \right] \non
& \phantom{aaa} - C_{ijk}C_{jlm} \left
  ( p_0^{(0)l}\del{2}p_0^{(0)m}\del{2}x_0^{(0)k} -
  p_0^{(0)l}\del{2}x_0^{(0)m} p_0^{(0)k} \right) \cdot \sigma_1 , \\
  \phi_2^i 
& = \del{1}p_0^{(2)i} - C_{ijk} \left[ p_0^{(1)j}\del{2}p_0^{(0)k} +
  p_0^{(0)j}\del{2}p_0^{(1)k} \right] \non
& \phantom{aaa} - C_{ijk}C_{jlm} \left
  ( p_0^{(0)l}\del{2}p_0^{(0)m}\del{2}p_0^{(0)k} -
  p_0^{(0)l}\del{2}p_0^{(0)m} p_0^{(0)k} \right) \cdot \sigma_1 ,
\end{align}
and we find the solutions,
\begin{align}
  x_0^{(2)i}(\tau,\sigma_1,\sigma_2) &=
    A_0^{(2)i}(\tau,\sigma_2) + C_{ijk} \left
    [ B_0^{(1)j}\del{2}x_0^{(0)k} + p_0^{(0)j}\del{2}A_0^{(1)k}
    \right] \sigma_1 \non
& \phantom{aaa} + C_{ijk}C_{jlm} \left
  ( p_0^{(0)l}\del{2}p_0^{(0)m}\del{2}x_0^{(0)k} -
  p_0^{(0)l}\del{2}x_0^{(0)m} p_0^{(0)k} \right) \cdot \frac{\sigma_1^2}{2}
 , \non
  p_0^{(2)i}(\tau,\sigma_1,\sigma_2) &=
    B_0^{(2)i}(\tau,\sigma_2) +C_{ijk} \left
    [ B_0^{(1)j}\del{2}p_0^{(0)k} + p_0^{(0)j}\del{2}B_0^{(1)k}
    \right] \sigma_1 \non
& \phantom{aaa} + C_{ijk}C_{jlm} \left
  ( p_0^{(0)l}\del{2}p_0^{(0)m}\del{2}p_0^{(0)k} -
  p_0^{(0)l}\del{2}p_0^{(0)m} p_0^{(0)k} \right) \cdot \frac{\sigma_1^2}{2}
  \ .
\end{align}
Putting them together, we find that the 
$X^i(\tau,\sigma_1,\sigma_2)$ and $P^i(\tau,\sigma_1,\sigma_2)$ are
determined by the unconstrained boundary values,
$X_0(\tau,\sigma_2) = x_0^{(0)} + A_0^{(1)} + A_0^{(2)}$ and 
$P_0(\tau,\sigma_2) = p_0^{(0)} + B_0^{(1)} + B_0^{(2)}$ as follows,
\begin{align}
  \label{perturb sol of X}
  X^i(\tau,\sigma_1,\sigma_2)
  = & X^i_0 + \sigma_1 C_{ijk}P_0^j \del{2} X_0^k \nonumber\\ 
       &  + \frac{\sigma_1^2}{2} C_{ijk}C_{jlm} \left[ 
        \del{2} X_0^k P_0 ^l \del{2} P_0^m - P_0^k \del{2} (P_0^l
      \del{2} X_0^m ) \right] , \\
  \label{perturb sol of P}
 P^i(\tau,\sigma_1,\sigma_2)
 = & P_0^i + \sigma_1 C_{ijk}P_0^j \del{2} P_0^k \nonumber\\
       & + \frac{\sigma_1^2}{2} C_{ijk}C_{jlm} \left[ 
        \del{2} P_0^k P_0 ^l \del{2} P_0^m - P_0^k \del{2} (P_0^l
      \del{2} P_0^m ) \right] .
\end{align}
One can confirm that these solutions satisfy the remaining constraints by
substituting (\ref{perturb sol of X}) and (\ref{perturb sol of P})
 into the explicit 
form of $\phi_3^i$ and taking into account the fact 
that the other higher constraints involve
only higher derivative terms of $\sigma_1$ and $\sigma_2$.
Since we get the solutions of the constraints, we can compute the Dirac
brackets of $X$ and $P$ by the method given in appendix \ref{app:Dirac}. 
 This is what we shall do in the following section.

\section{Computing the Dirac brackets}
\label{sec:comp}

In order to compute the Dirac brackets,
 we first calculate Lagrange brackets.
In this case, Lagrange bracket $\textbf{L}$ is defined as
\begin{align}
  \label{symp form}
  \Omega = & -2 \int\!\! d^2 \sigma dX^i(\sigma_1,\sigma_2) 
            \wedge dP^i(\sigma_1,\sigma_2) \nonumber\\
          = & \int\!\! dxdy \, {\bf L}^{ij}_{xy} \, d\phi^i(x)
          \wedge d\phi^j(y) , 
\end{align}
where we have integrated over $\sigma_1$,
$d\phi = dX_0(\sigma_2)$ or $dP_0(\sigma_2)$, and $x$ and  $y$ denote
the $\sigma_2$ coordinate.
Dirac bracket $\textbf{C}$ is determined by the inverse matrix 
of this Lagrange brackets, $\textbf{C} = \textbf{L}^{-1}$.
To calculate the Lagrange bracket of this system, we determine the
effects of the $C$-field order by order, to order $C^2$:
\begin{equation}
  {\bf L} =  {\bf L}^{(0)} + {\bf L}^{(1)} + {\bf L}^{(2)} ,
\end{equation}
where $L^{(i)} $ denotes the terms of order $C^i$.
Then the Dirac bracket is obtained as
\begin{align}
\label{inverseandC}
  {\bf C} = \textbf{L}^{-1} 
= & {\bf L}^{(0) { -1}} - {\bf L}^{(0) { -1}}
  ( {\bf L}^{(1)}
  + {\bf L}^{(2)} ) {\bf L}^{(0) { -1}} + {\bf L}^{(0) { -1}}{\bf
  L}^{(1)}{\bf L}^{(0) { -1}}{\bf L}^{(1)}{\bf L}^{(0) {
  -1}} + {\cal O}(C^3) \\
          = & {\bf J} - {\bf J} ( {\bf L}^{(1)} + {\bf L}^{(2)} )
  {\bf J} + {\bf J}{\bf L}^{(1)}{\bf J}{\bf L}^{(1)}{\bf J} + {\cal
  O}(C^3) ,
\end{align}
where we have abbreviated $ {\bf  L}^{(0) { -1}}$ as ${\bf J}$.

Let us start the calculation. In zeroth order in $C$, 
the Lagrange bracket is determined through the symplectic form
\begin{align}
  \Omega^{[0]} = & -2 \int\!\! d\sigma^2 dX_0^i \wedge dP_0^i
  \nonumber\\
  = & -2\pi L \int\!\! dxdy \, \delta^{ij} \, \delta(x-y) \, dX_0^i(x) \wedge
  dP_0^j(y) \ .
\end{align}
We get
\begin{equation}
  {\bf L}^{(0)} = \left(
    \begin{array}{cc}
0 & L^{(0)} \\
-\trs{(L^{(0)})} & 0 \\
    \end{array}
\right) , 
\end{equation}
where
\begin{equation}
  L^{(0)} = -\pi L \delta^{ij}\delta(x-y) .
\end{equation}
The inverse matrix of this ${\bf L}^{(0)}$ is
given by
\begin{equation}
{\bf J}= ({\bf L}^{(0)})^{-1} = \left(
  \begin{array}{cc}
 & -J \\
J & \\
  \end{array}
\right) \ , \ J =( L^{(0)})^{-1} = - \frac{1}{\pi L} \, \delta^{ij} \,
\delta(x-y)
\ , \ \trs{J} = J \ .
\end{equation}
At this stage, we can calculate the Dirac bracket at $C=0$:
\begin{align}
  \left\{X_0^i(x) , X_0^j(y) \right\}_{\rm DB} = & 0, \\
  \left\{P_0^i(x) , P_0^j(y) \right\}_{\rm DB} = & 0, \\
  \left\{X_0^i(x) , P_0^j(y) \right\}_{\rm DB} = & \frac{1}{\pi L} \,
  \delta^{ij} \, \delta(x-y)   \ .
\end{align}
These are the original Poisson brackets except for
the normalization factor.

\paragraph{Calculations of ${\cal O}(C^1)$}

Next, we shall calculate the $C^1$ part.
This is the first non-trivial result in 
these calculations.
The symplectic form of this order is
\begin{align}
\label{Omega1}
  \Omega^{[1]} = & -2 \int\!\! \, d^2\sigma \ \left[ \sigma_1 \, C_{ikl} \,
  dX_0^i \wedge ( dP_0^k \del{2} P_0^l + P_0^k \del{2} dP_0^l ) \right.
  \nonumber\\ 
  & \phantom{ -2 \int\!\! d^2\sigma \sigma_1 C_{ikl}}\left. +
  \sigma_1 C_{ikl}( dP_0^k \del{2} X_0^l + P_0^k \del{2} dX_0^l )
  \wedge dP_0^i \right] \non
  = & -  (\pi L)^2 \int\!\!dxdy \, C_{ijl} \, \left[ dX_0^i(x) \wedge dP_0^j(y)
    \left( -2 C_{ijl} P_0^l(x) \del{x} \delta(x-y) \right) \right.
  \non
  & \phantom{aaaaaaaaaaaaaaaaa} \left. - dP_0^i(x) \wedge dP_0^j(y)
  \del{x}X_0^l \, \delta(x-y) \right] ,
\end{align}
and we get
\begin{equation}
  {\bf L}^{(1)}  =  \left(
    \begin{array}{cc}
0 & L^{(1)} \\
-\trs{(L^{(1)})} & l^{(1)} \\
    \end{array}
\right) ,
\end{equation}
where
\begin{align}
 L^{(1)} = &  (\pi L)^2 C_{ijl} P_0^l(x) \del{x} \delta(x-y) , \\
l^{(1)}  = & (\pi L)^2 C_{ijl} \del{x} X_0^l \delta(x-y) \ .
\end{align}

At this order, the Dirac bracket is
\begin{align}
  \left\{X_0^i(x) , X_0^j(y) \right\}_{\rm DB} = & C_{ijl} \del{x}
  X_0^l \, \delta(x-y) , \\
  \left\{P_0^i(x) , P_0^j(y) \right\}_{\rm DB} = & 0 ,\\
  \left\{X_0^i(x) , P_0^j(y) \right\}_{\rm DB} = &
  \frac{1}{\pi L} \,  \delta^{ij} \, \delta(x-y) -
  C_{ijl} P_0^l(y) \deldash (y-x)  \label{XP1order} \ .
\end{align}
One can check that the Jacobi identity holds at this order, 
\begin{align}
 &    \left\{ \left\{ X_0^i(x) , P_0^j(y) \right\} , X_0^k(z) \right\} + 
    \mbox{\rm (cyclic.) }  \non
 & =  \frac{1}{\pi L} C_{ijk} \left( \delta(y-z) \deldash (y-x) +
    \delta(y-x) \deldash (y-z) + \delta(z-x) \deldash (z-y)  \right)
    \non
 & =  \frac{1}{\pi L} C_{ijk} \left( \delta(y-z) \deldash (y-x) +
    \delta(y-x) \deldash (y-z) + \delta(y-x) \deldash (z-y) -
   \deldash (z-x) \delta(z-y) \right) \non
 & = 0 \ .
\end{align}
The Jacobi identity for $\PB{X}{\PB{X}{X}}$ is
trivially satisfied at first order in $C$.
To see how it is non-trivially satisfied, we
turn to the calculations of $C^2$.

\paragraph{Calculations of ${\cal O}(C^2)$}

The calculations of order $C^2$ turn out to be very complicated, so we split
the calculations into some parts. 

First, we consider the cross terms, 
$(C^1 \text{ part}) \wedge (C^1\text{ part})$ . 
The symplectic form of this part is
\begin{align}
\label{Omega2-1}
  \Omega^{[2-1]}  = & -2 \int\!\! d^2\sigma \, \sigma_1^2 \ C_{ijk}C_{ilm} \,
  ( dP_0^j \del{2}X_0^k + P_0^j  \del{2} dX_0^k) \wedge ( dP_0^l
  \del{2}P_0^m + P_0^l \del{2} dP_0^m) \non
  = & -\frac{2(\pi L)^3}{3} \int\!\! d^2\sigma \, C_{ikl}C_{jml} \non
  & \times \left\{ dX_0^i(x) \wedge dP_0^j(y) \del{x} \left
  ( P_0^k(x) \left(  2\del{y} P_0^m(y) + P_0^m(y) \del{y} \right)
  \delta(x-y) \right) \phantom{\frac{}{}} \right. \non
 & \phantom{aaaaaa}  + dP_0^i(x) \wedge dP_0^j(y) \left[ 
 \frac{1}{2} \left( \Xd{k} (x) \Pd{m}(x) - \Pd{k}(y) \Xd{m}(y) \right)
  \delta(x-y) \right. \non
  & \left. \left. \phantom{\frac{}{}AAAAAAAAAAAAAA}
 - \left( \Xd{k}(x)P_0^m(x) + P_0^k(y)
  \Xd{m}(y) \right) \deldash (x-y)  \right]  \right\} \ ,
\end{align}
so we get
\begin{equation}
  {\bf L}^{[2-1]} = \left(
    \begin{array}{cc}
 & L^{[2-1]} \\
-\trs{(L^{[2-1]})} & l^{[2-1]} \\
    \end{array}
\right) \ ,
\end{equation}
where
\begin{align}
 L^{[2-1]} & =   - \frac{(\pi L)^3}{3} C_{ikl}C_{jml} \del{x} \left
  ( P_0^k(x) \left(  2\del{y} P_0^m(y) + P_0^m(y) \del{y} \right)
  \delta(x-y) \right) , \\
  l^{[2-1]} & = 
   -\frac{1}{3}(\pi L)^3 C_{ikl}C_{jml} \left( \left( \Xd{k} (x)
  \Pd{m}(x) - \Pd{k}(y) \Xd{m}(y) \right) \delta(x-y) \right. \non
  &   \phantom{AAAAAAA} - \left. \left( \Xd{k}(x)P_0^m(x) + P_0^k(y)
  \Xd{m}(y) \right) \deldash (x-y) \right) \ . 
\end{align}

Next, we consider the $(C^0 \text{ part}) \wedge (C^2\text{ part})$.
The symplectic form of this part is
\begin{align}
  \label{Omega2-2}
  \Omega^{[2-2]} &= -2 \int\!\! d^2\sigma \, \sigma_1^2 \
  C_{ijk}C_{ilm} \non
& \phantom{aaaa} \times \left\{ \left[  \del{2} dX_0^k P_0^l  \del{2} P_0^m + \del{2}X_0^k dP_0^l
 \del{2} P_0^m  - \del{2}X_0^k P_0^m \del{2} dP_0^l
  \right. \right. \nonumber\\
  & \phantom{aaaaaaaaaaaaa} \left. - dP_0^k \del{2}(P_0^l \del{2}
  X_0^m) - P_0^k \del{2}(dP_0^l \del{2} X_0^m - P_0^m \del{2}d X_0^l)
 \right]  \wedge dP_0^i \non
 & \phantom{aaaaaaa} + dX_0^i \wedge \left[ \del{2} dP_0^k P_0^l
  \del{2} P_0^m - \del{2}P_0^k  P_0^m \del{2} dP_0^l 
  + \del{2}P_0^k dP_0^l \del{2} P_0^m \right. \nonumber\\
  & \left. \left. \phantom{aaaaaaaaaaaaaa}  - dP_0^k \del{2}(P_0^l \del{2}
  P_0^m) - P_0^k \del{2}(dP_0^l \del{2} P_0^m - P_0^m \del{2}d P_0^l)
  \right] \right\} .
\end{align}
Then we find that the $\Omega^{[2-2]}$ has the form
\begin{equation}
\Omega^{[2-2]} = \int\!\! dxdy {\bf L}^{[2-2]} d\phi^i(x) \wedge
d\phi^j(y)  \ ,
\end{equation}
where
\begin{equation}
  {\bf L}^{[2-2]} = {\bf M} + {\bf N} ,
\end{equation}
\begin{equation}
  ({\bf M})^{ij}_{xy} = \left(
    \begin{array}{cc}
 & M \\
-\trs{M} & m \\
    \end{array}
\right) \ ,
\end{equation}
\begin{equation}
  ({\bf N})^{ij}_{xy} = \left(
    \begin{array}{cc}
 & N \\
-\trs{N} & n \\
    \end{array}
\right) \ ,
\end{equation}
and, ${\bf M}$ and ${\bf N}$ correspond to the following tensor structures of
 $C^2$:
\begin{eqnarray*}
  {\bf M} & \propto & C_{ijk}C_{klm}  , \\
  {\bf N} & \propto & C_{ikl}C_{jml}  \ .
\end{eqnarray*}
The explicit calculations of $\textbf{M}$ and $\textbf{N}$
are shown in appendix \ref{app:calc}.
The results are
\begin{align}
\label{resultofM}
  M & = \frac{(\pi L)^3}{3} C_{ijk}C_{klm} \left[ P_0^l(x) \del{x}
  P_0^m(x) \deldash (x-y) \right] , \\
\label{resultofm}
  m & = \frac{(\pi L)^3}{3} C_{ijk}C_{klm} \del{y} \left( P_0^l(y)
  \del{y}X_0^m(y) \right) \delta(x-y) , \\
\label{resultofN}
  N & =  \frac{(\pi L)^3}{3} C_{ikl}C_{jml} \left[P_0^k(x)P_0^m(x)
  \delta^{\prime \prime} (x-y) +  \Pd{m}(x) \del{x} \left( P_0^k(x)
  \delta (x-y) \right) \right]  , \\
\label{resultofn}
 n & =  \frac{(\pi L)^3}{3} C_{ikl}C_{jml} \left[ \Xd{k}(x)P_0^m(x) +
  \Xd{m}(y)P_0^k(y) \right] \deldash (x-y) .
\end{align}
Thus we get the Lagrange brackets to order $C^2$.
Let us compute the Dirac brackets.

\paragraph{Computing the Dirac brackets}

By (\ref{inverseandC}), we can calculate the Dirac brackets
$\textbf{C}$,
\begin{equation}
  \textbf{C}^{ij}_{xy} =
  \begin{pmatrix}
     \DB{X_0^i(x)}{ X_0^j(y)} & \DB{X_0^i(x)}{P_0^j(y)} \cr
     - \DB{P_0^j(y)}{X_0^i(x)} & \DB{P_0^i(x)}{P_0^j(y)} \cr
  \end{pmatrix} ,
\end{equation}
as follows:
\begin{align}
   \DB{X_0^i(x)}{ X_0^j(y)} & =
  J \left( l^{(1)}\right)J  + J \left( l^{(2)} \right) J -
  Jl^{\small(1)} J L^{\small (1)} J -  J\trs{(L^{\small(1)})} J
  l^{\small (1)} J \non
 & = \frac{1}{(\pi L)^2} \left( l^{(1)} \right)^{ij}_{xy} 
 +\frac{1}{(\pi L)^2} \left( l^{(2)} \right)^{ij}_{xy} \non
& \ph{aaa} + \frac{1}{(\pi L)^3} \left\{ (l^{\small(1)})^{il}_{xz}(L^{\small
  (1)})^{lj}_{zy} 
+ \left( \trs{(L^{\small(1)})} \right)^{il}_{xz} (l^{\small
  (1)})^{lj}_{zy} \right\} + {\cal O}(C^3)  , \\
  \DB{X_0^i(x)}{ P_0^j(y)} & = -J + J \left( \trs{(L^{(1)})} \right)J
  + J \left( \trs{(L^{(2)})} \right) J - J
  \trs{(L^{\small(1)})} J \trs{(L^{\small (1)})} J \non & =
  \frac{1}{\pi L}({\bf 1})^{ij}_{xy} + \frac{1}{(\pi L)^2}
  \left( \trs{(L^{(1)})} \right)^{ij}_{xy} \non & \ph{aaa}
  +\frac{1}{(\pi L)^2} \left( \trs{(L^{(2)})} \right)^{ij}_{xy} +
  \frac{1}{(\pi L)^3} \left( \trs{(L^{\small(1)})} \right)^{il}_{xz} \left(
    \trs{(L^{\small(1)})} \right)^{lj}_{zy} + {\cal O}(C^3) , \\
\DB{P_0^i(x)}{P_0^j(y)} &= 0 \ .
\end{align}
Explicit computation shows
\begin{align}
  \DB{X_0^i(x)}{ X_0^j(y)} &=
 C_{ijl} \Xd{l}(x) \delta(x-y) \non
&\ph{aa}
 - \frac{1}{3}C_{ikl}C_{jml} \left[ \left( \Xd{k}(x) \Pd{m}(x)
   - \Xd{m}(y) \Pd{k}(y) \right) \delta (x-y) \right. \non
 & \left. \phantom{aaAAAAAAAAA }
 +  \left( \Xd{k}(x) P_0^m(x)
   + \Xd{m}(y) P_0^k(y)  \right) \deldash (x-y) \right] \non
  &\ph{aaaa}  
  + \frac{1}{3} C_{ijk}C_{klm} \del{y} \left( P_0^l(y) \Xd{m}(y) 
  \right) \delta(x-y) \ +{\cal O}(C^3), \label{resultofXX} \\
  \DB{X_0^i(x)}{P_0^j(y)} & =  \delta^{ij} \delta(x-y)
+   C_{ijl} P_0^l(y) \deldash (x-y) \non
    &\ph{aa}
 -\frac{1}{3}C_{ikl}C_{jml} \left[ P_0^k(x)P_0^m(x)
  \delta^{\prime \prime} (x-y) \right. 
  + 3P_0^k(x)\Pd{m}(x)  \deldash (x-y) \non
  &  \phantom{aa-\frac{\pi L}{3}C_{ikl}C_{jml} a} \left. + \left
  ( 2P_0^k(x)P_0^{\prime\prime m}(x) + \Pd{k}(x) \Pd{m}(x) \right)
  \delta(x-y) \right] \non
&\ph{aaaa}
 + \frac{1}{3} C_{ijl}C_{lkm} P_0^k(y)\Pd{m}(y) \deldash (x-y) 
\ + {\cal O}(C^3) \label{resultofXP} \ ,
\end{align}
where we have rescaled the momenta, $\pi L P_0^i \rightarrow P^i_0$.
This is because in the limit $L \rightarrow 0$, 
the integrated momenta $\pi LP_0$ are more naturally assigned to the
boundary strings than the original boundary momenta $P_0$.

These results mean that the coordinates of the boundary strings of 
an open membrane
in the constant $C$-field background show \textit{noncommutativity}. 
It is very curious that the commutation relation between $X^i$ and $X^j$ 
depends on other components of transverse fields, $X^k$.

\section{Concluding remarks}
\label{sec:conc}

In the previous section, we have obtained  
the Dirac brackets of an open membrane in 
the $C$-field background.
The result  shows that the boundary string has a loop-space
noncommutativity.

We can confirm that the Jacobi identity holds at order in $C^2$ with these
results, though we do not write down the calculation explicitly.
Indeed, the satisfaction of Jacobi identity is trivial from the general
properties of Poisson bracket, but the cancellations between the terms are not
trivial. This indicates the algebra has complicated structures and
more transparent understanding of it from the boundary string
viewpoint is desirable.

The results presented above are the Dirac brackets between the
coordinates and momenta of the boundary.
Dirac brackets between the coordinates on the membrane can be calculated by 
(\ref{perturb sol of X}),
and there exists noncommutativity not only at the
boundary but also on the membrane.
In string theory, the string coordinates are commutative except at
its ends as explained in appendix \ref{app:Dirac}, and to show this
it is essential to include all the oscillation modes.
Thus we also expect that including all the oscillation modes 
make the membrane coordinates commutative except at its boundary,
because the $C$-field part of the action (\ref{C-part}) is total
derivative for a constant $C$,
and should change the dynamics only at the boundary.

We have done our analysis in a tractable static gauge condition.
Light-cone gauge analysis is more interesting in its relationship with
BFSS matrix theory and the results of \cite{CDS97}.
It is easy to find the light-cone gauge Hamiltonian,
\begin{equation}
  H_{\textrm{LC}} = \int \!\! d^2 \sigma \frac{1}{2P^+} 
\left[ (P^i + C_{ijk} \del{1}X^j \del{2} X^k)^2 + \frac{T^2}{2}
  \PB{X^i}{X^j}^2 \right] ,
\end{equation}
the equations of motion
\begin{equation}
  \ddot{X}^i + \PB{X^j}{\PB{X^i}{X^j}} = 0 ,
\end{equation}
and the boundary conditions
\begin{equation}
\left.  -T \del{2}X^j \PB{X^i}{X^j} + C_{ijk} \del{2}X^j \dot{X}^k
  \right|_{\sigma_1 = 0 , \pi } = 0 .
\end{equation}
However, the chain of the boundary constraints look too complicated to solve
in this case even if some approximations are taken.
Moreover, when there is a constant $C$-field background,
we can not apply the matrix
regularization method developed in the third paper of \cite{openMem}.
Thus, analysis in this gauge is remaining as a hard but interesting
problem.
See comments below.

\vspace*{1.5cm}

When this work was in the process of typing, we learned that another group
\cite{BBSS00} has also employed the quantization of an open membrane in
a $C$-field background, and they have also investigated the decoupling limit
as the open string case. 
Though their line of thought is different from ours, 
their results seem to be consistent with ours at least in first
order in $C$.
Moreover, their paper has also studied the light-cone coordinate analysis,
but their analysis is within the decoupling limit and slightly
different from our interests such as membrane regularization
related to matrix models.

\section*{Acknowledgments}

S.~K. is supported in part by the Japan Society for the
Promotion of Science under the Predoctoral Research Program.
N.~S. is supported in part by Grant-in-Aid for Scientific Research
(\#12740150), and in part by Priority Area: 
``Supersymmetry and Unified Theory of Elementary Particles'' (\#707), 
from Ministry of Education, Science, Sports and Culture. 

\appendix

\section{A brief review of Dirac's procedure applied to boundary constraints}
\label{app:Dirac}

In string theory, one can find noncommutativity on a D-brane by
quantization procedures for open strings with a background
$B$-field \cite{CH98}.
A transparent way to confirm the noncommutativity of open strings is the
Dirac's procedure applied to boundary conditions \cite{CH99,AAS-J99}.
In this appendix, we briefly review this approach.
The calculations described here are mainly based on the appendix of
the paper by Kawano and Takahashi 
\cite{KT00}.

\paragraph{Dirac's procedure}

First, we survey the ordinary methods for constrained systems following 
\cite{Diractext,HRTtext}.
In singular systems, we face some constraints,
\textit{primary constraints}, between canonical variables.
Consistency conditions for these constraints in time evolution
sometimes lead to additional constraints, \textit{secondary constraints}.
We must consider the consistency conditions for these new constraints and
possibly find new constraints, secondary constraints for secondary
constraints, and so on.

Constraints are classified into two classes;
the first class constraints that commute with  all the other 
constraints and the second class constraints that do not.
The first class constraints are related to the gauge symmetry of the system
and we can treat them as second class by gauge fixing.
Thus we may assume all the constraints are second class.
The singular system is treated with the Dirac bracket defined as
\begin{equation}
  \DB{F}{G} \equiv \PB{F}{G}  - \PB{F}{\phi_A}C^{AB}\PB{\phi_B}{G},
\end{equation}
where $C^{AB} = (C^{-1})^{AB}$,  
$C_{AB} \equiv \PB{\phi_A}{\phi_B}$ and $\phi_A$, $\phi_B$ are second
class constraints. 
This Dirac brackets are Poisson brackets on the constrained
surface\cite{Diractext,HRTtext}, so we can determine the time
evolution of this constrained system using the Dirac bracket.

\paragraph{Boundary condition as constraint}

According to \cite{BCasDC}, we can treat the boundary conditions of an 
open string as constraints.
The consistency conditions of these
constraints lead to an infinite chain of secondary constraints, 
which are all second class.
Thus, we can calculate Dirac brackets of this system in principle.
However, we must consider the inverse of an $\infty \times \infty$
matrix $C_{AB}$.
Surprisingly, we can completely solve this question in the string case.

Let us explain the string case calculations for example.
We consider an open string in a constant NS-NS $B$-field background. 
The action of this system is
\begin{equation}
  S = \frac{1}{4\pi \alpha'} \int_{\Sigma} \!\! d^2 \sigma \left[
   g_{ij} \left( \dot{X}^i \dot{X}^j - \Xp{i} \Xp{j} \right) + 2b_{ij}
  \dot{X}^i\Xp{j} \right] \ ,
\end{equation}
where 
\begin{equation}
  X' \equiv \frac{\partial}{\partial \sigma} X \ \ , \ \ \dot{X}
  \equiv \frac{\partial}{\partial \tau} X \ ,
\end{equation}
and $b_{ij} = 2 \pi \alpha' B_{ij}$.
Variation of the action leads to the equations of motion and the boundary
conditions:
\begin{equation}
  \partial^{\alpha} \partial_{\alpha} X^i(\tau , \sigma ) =0 ,
\end{equation}
\begin{align}
  \mbox{Dirichlet directions:}\quad & \delta X^{i_{{\rm D}}} =0
 \ \ \ \  (X^{i_{{\rm  D}}}= \mbox{const.})  , \non
  \mbox{Neumann (or Mixed) directions:}\quad & g_{ij} X^{\prime j} +
  b_{ij}\dot{X}^j =0 \ \ \mbox{at} \ \ \sigma  = 0, \pi \nonumber ,
\end{align}
where
mixed directions are named for their mixtures of some
directions\cite{BCasDC} and 
we shall only consider below the directions obeying 
these mixed boundary conditions. 
We now go on to the canonical formalism. Conjugate momenta are 
$2 \pi \alpha' P_i (\tau,\sigma ) =  \left( g_{ij}\dot{X}^j + b_{ij} \Xp{j}
  \right)$ and the boundary conditions are taken to be 
\textit{primary constraints} of this system,
\begin{equation}
  \label{def of phi}
  \phi_i(\sigma) = G_{ij} \Xp{j}  + 2\pi  \alpha' b_{ik}g^{kl}P_l ,
\end{equation}
where $G_{ij} \equiv g_{ij} - (bg^{-1}b)_{ij}$, so called ``open
string metric''.

The consistency of the constraints in time evolution leads to 
an infinite chain of secondary constraints:
\begin{equation}
  \frac{\partial^{(2n+1)}}{\partial \sigma^{(2n+1)}} P_i(\sigma) \we 0
  \quad \mbox{and} \quad 
  \frac{d^{(2n)}}{d\sigma^{(2n)}} \phi_i(\sigma) \we 0 \ .
\end{equation}
The solution to these constraints is\cite{KT00}
\begin{align}
  \label{expansion of X and P} 
X^i(\tau,\sigma) &= \sum_{n=0}^{\infty} X_n^i(\tau)\cos (n\sigma) +
\Theta^{ij} \left[ P_{0j}(\tau)\sigma + \sum^{\infty}_{n=1}
  \frac{1}{n}P_{nj} \sin (n\sigma) \right]  , \\ 
P_i(\tau,\sigma) &= 
 \sum^{\infty}_{n=0} P_{ni}(\tau) \cos (n\sigma) .
\end{align}

\paragraph{Lagrange bracket}

One of the easiest way to find the Dirac bracket is to use the
Lagrange brackets\cite{HRTtext} and this method was used in
\cite{CH98} in a slightly different way.

Lagrange bracket \textbf{L} for variables $ z^\mu = z^\mu(q,p)$ 
is defined through the symplectic form
\begin{equation}
  \Omega = -2 dq^i(z) \wedge dp_i(z) 
= {\bf L}^{\mu\nu} dz^\mu \wedge dz^\nu \ ,
\end{equation}
where $q$ and $p$ are canonical variables of this system.
Explicitly, Lagrange bracket is written as
\begin{equation}
  \label{def of Lagrange bracket}
 {\bf L}^{\mu\nu} = \frac{\partial q^i}{\partial z^\mu}
\frac{\partial p_i}{\partial  z^\nu } - 
\frac{\partial q^i}{\partial z^\nu}\frac{\partial p_i}{\partial z^\mu} 
\ .
\end{equation}
An important property of this bracket is that this is the inverse matrix
of the Poisson bracket,
\begin{equation}
  {\bf L}_{\mu\nu} \PB{z^\nu}{z^\rho} = \delta^\rho_\mu \ .
\end{equation}

To find the relation to the Dirac bracket, let us take 
the variables as follows
\begin{equation}
  \label{choice}
  \underbrace{z^1,z^2, \ldots , z^{2N-2m}}_{\text{coodinates on the
      constrained surface}} , 
\underbrace{z^{2N-2m+1} = \phi_1 , \ldots , z^{2N} 
  = \phi_{2m}}_{\text{2m constraints}} \ .
\end{equation}
Then we find that the matrix obtained by limiting variables to
the first $(2N-2m)$ ones is the inverse matrix of the Dirac bracket,
\begin{equation}
  \sum_{\mu,\nu=1}^{2N-2m} {\bf L}_{\mu\nu} \DB{z^\nu}{z^\rho}
=  \delta^\rho_\mu \ .
\end{equation}
This means that Dirac bracket is the Poisson bracket on the constrained
surface defined through the conditions, $z^{2N-2m+1} =\ldots = z^{2N}=0$.
Thus, we can compute the Dirac bracket by solving the constraints, 
constructing the Lagrange bracket and taking its inverse.

In string case, Lagrange brackets are defined by
\begin{align}
\label{stringsymp}
  \Omega &= -2 \int\!\! d\sigma dX^i(\sigma) \wedge dP_i(\sigma) \non
     & = -2 \left[\pi dX^i_0 \wedge dP_{0i} + \frac{\pi}{2} dX^i_n
  \wedge dP_{ni} - \Theta^{ij} \frac{\pi^2}{2} dP_{0i} \wedge dP_{0j}
  \right] \ .
\end{align}
{}From this, we can determine 
the Lagrange brackets for every mode of $X$ and $P$.
Taking the inverse, we obtain
\begin{align}
  \DB{X^i(\sigma)}{P_j(\sigma')} &= \delta^i_j \left( 
\frac{1}{\pi} + \frac{2}{\pi} \sum_{n=1}^\infty \cos (n\sigma) \cos
(n\sigma') \right) \non
&\equiv \delta^i_j \; \tilde{\delta}(\sigma,\sigma') \label{delchilda} \\
\DB{P_i(\sigma)}{P_j(\sigma')} &= 0 \\
\DB{X^i(\sigma)}{X^j(\sigma')}
&= \left\{
  \begin{array}{ll}
\Theta^{ij} & (\sigma=\sigma'=0) \\
-\Theta^{ij} & (\sigma=\sigma'=\pi) \\
0 & (\text{otherwise}) \\
  \end{array} \right. \ .
\end{align}
This shows noncommutativity of open strings and this equals
the result in \cite{CH98}.

\section{The explicit calculations of Lagrange brackets at second order
  in $C$}
\label{app:calc}

In this appendix, we give the explicit calculations 
of (\ref{resultofM}), (\ref{resultofm}), (\ref{resultofN}) and
(\ref{resultofn}).

First, we calculate the part of $M$.
This part of the symplectic form is
\begin{align}
  \Omega^{[2-2]}_M &= -\frac{(\pi L)^3}{3} \int \!\! dxdy C_{ijk}C_{jlm}
  \non
& \mbox{} \times \left[ dX_0^i(x) \wedge dP_0^k(y) \left[
-\del{y}\left( P_0^l(y)\Pd{m}(y) \delta(x-y) \right)
  -\del{y}(P_0^l(y)\Pd{m}(y))\delta(x-y) \right] \right. \non
& \phantom{aaaaaaaaaa} \left. + dX_0^k(x) \wedge dP_0^i(y) \left[
-\del{x} \left( P_0^l(x) \Pd{m}(x)\delta(x-y) \right) \right] 
\right] \non
&= \frac{2(\pi L)^3}{3} \int \!\! dxdy C_{ijk}C_{jlm}
 dX_0^i(x) \wedge dP_0^j(y) \left[
 P_0^l(x)\Pd{m}(x) \deldash (x-y) \right] .
\end{align}
These correspond to $(2M)^{ij}_{xy} dX_0^i(x) \wedge dP_0^j(y)$, 
and hence
\begin{equation}
  M = \frac{(\pi L)^3}{3} C_{ijk}C_{klm} \left[ P_0^l(x) \del{x}
  P_0^m(x) \deldash (x-y) \right] .
\end{equation}

Next, we consider the $N$ part.
\begin{align}
  \Omega^{[2-2]}_N &= -\frac{(\pi L)^3}{3} \int \!\! dxdy C_{ilk}C_{ljm} \non
& \phantom{aa} \times \left\{ dX_0^i(x) \wedge dP_0^j(y) \left[
 \Pd{m}(y) \Pd{k}(y) \delta(x-y) + \del{y}\left( P_0^m(y)\Pd{k}(y)
  \delta(x-y) \right) \right. \right. \non
& \phantom{aaaaaa} \left. 
+ \Pd{m}(y) \del{y}\left(P_0^k(y)\delta(x-y) \right) 
+\del{y} \left(P_0^m(y) \del{y} \left( P_0^k(y) \delta(x-y) \right) 
\right) \right] \non
& \phantom{aaaaaaaaa} \left.
+ dX_0^j(x) \wedge dP_0^i(y) \left[ 
\del{x} \left( P_0^m(x) \del{x} \left( P_0^k(x) \delta(x-y) \right)
  \right) \right] \right\} \non
&= \frac{(\pi L)^3}{3}C_{ikl}C_{jml} \int \!\! dxdy \, dX_0^i(x) \wedge
  dP_0^j(y) \non
& \phantom{aa} \times \left[
\Pd{m}(x) \Pd{k}(x) \delta(x-y) - P_0^m(x)\Pd{k}(x) \deldash (x-y) 
\right. \non
& \phantom{aaaaaaaaa} 
-\Pd{m}(y)P_0^k(x)\deldash (x-y) 
- \del{y} \left( P_0^m(y) P-)^k(x) \deldash (x-y) \right) \non
& \phantom{aaaaaaaaa} \left.
+ \del{x} \left( P_0^k(x) \del{x} \left( P_0^m(x) \delta(x-y) \right) 
\right) \right] ,
\label{calcN1}
\end{align}
where in the last term we make $k \leftrightarrow m$.
Using
\begin{align}
  -P_0^k(x)\Pd{m}(y) \deldash (x-y) &= -P_0^k(x) \left(P_0^{\prime
  \prime m}(x) \delta(x-y) + \Pd{m}(x) \deldash (x-y) \right)  , \non
-\del{y} \left(P_0^k(x)P_0^m(y)\deldash (x-y) \right) &=
  P_0^k(x)\Pd{m}(x)\deldash (x-y) 
     + P_0^k(x)P_0^m(x) \delta^{\prime \prime} (x-y)  , \non
\del{x} \left( P_0^k(x) \del{x} \left( P_0^m(x) \delta(x-y) \right) 
\right) &=
\Pd{k}(x)\Pd{m}(x) \delta(x-y) + \Pd{k}(x)P_0^m(x) \deldash (x-y)  \non
& \phantom{aa} +P_0^k(x)P_0^{\prime \prime}(x)\delta(x-y) 
+2P_0^k(x)\Pd{m}(x) \deldash (x-y)   \non
& \phantom{aa} + P_0^k(x)P_0^m(x) \delta^{\prime\prime} (x-y),
\end{align}
we obtain
\begin{align}
(\text{\ref{calcN1}})
&=  \frac{(\pi L)^3}{3}C_{ikl}C_{jml} \int \!\! dxdy \, dX_0^i(x) \wedge
  dP_0^j(y)   \non
& \phantom{aa} \times 
\left[ 2P_0^k(x)P_0^m(x) \delta^{\prime\prime}(x-y) \right.  
+2P_0^k(x)\Pd{m}(x)\deldash (x-y)  \non
& \phantom{aaaaaaa}\left.
 +2\Pd{k}(x) \Pd{m} (x) \delta(x-y) \right] .
\end{align}
Hence,
\begin{align}
  N & =  \frac{(\pi L)^3}{3} C_{ikl}C_{jml} \left[P_0^k(x)P_0^m(x)
  \delta^{\prime \prime} (x-y) +  \Pd{m}(x) \del{x} \left( P_0^k(x)
  \delta (x-y) \right) \right] .
\end{align}

The $m$ part can be calculated as follows.
\begin{align}
  \Omega^{[2-2]}_m &= -\frac{(\pi L)^3}{3} \int \!\! dxdy C_{ijk}C_{jlm}
  dP_0^k(x) \wedge dP_0^i(y) \left( -\del{x} \left
  ( P_0^l(x)\Xd{m}(x) \right) \delta(x-y) \right) \non
& = \frac{(\pi L)^3}{3} \int \!\! dxdy C_{ijk}C_{klm}
   dP_0^i(x) \wedge dP_0^j(y) \left( -\del{x} \left
  ( P_0^l(x)\Xd{m}(x) \right) \delta(x-y) \right) ,
\end{align}
then
\begin{equation}
  m = \frac{(\pi L)^3}{3} C_{ijk}C_{klm} \del{y} \left( P_0^l(y)
  \del{y}X_0^m(y) \right) \delta(x-y) \ .
\end{equation}

Finally, we compute the part of $n$.
Because the result should be antisymmetric under 
$\{ i,x \} \leftrightarrow \{ j,y \}$, and $n$ is proportional 
to $C_{ikl}C_{jml}$, we only need to consider
 the antisymmetric part under  $\{ k,x \} \leftrightarrow 
\{ m,y \}$.
\begin{align}
  \Omega^{[2-2]}_n &= -\frac{(\pi L)^3}{3} \int \!\! dxdy C_{ijk}C_{jlm}
  \non
  & \mbox{} \times dP_0^l(x) \wedge dP_0^i(y) \left
  ( \Pd{m}(x)\Xd{k}(x) \delta (x-y) + \del{x}(P_0^m(x)\Xd{k}(x)\delta
  (x-y)) \right. \non
& \phantom{aaaaaaaaaaaaaaaaaaaaaaaaaaa} \left.+
  \Xd{m}(x)\del{x}(P_0^k(x)\delta (x-y)) \right) \non
  &= -\frac{(\pi L)^3}{3} \int \!\! dxdy C_{ijk}C_{jlm}
  \non
  & \mbox{} \times dP_0^l(x) \wedge dP_0^i(y) \left(
\Pd{m}(x)\Xd{k}(x) \delta (x-y) + \Pd{k}(x)\Xd{m}(x) \delta (x-y)
  \right. \non
& \phantom{aaaaaaaaaaaaaaaaaaa} \left. + \Xd{m}(x)P_0^k(x)  \deldash
  (x-y)  + \Xd{k}(y)P_0^m(y)  \deldash (x-y) \right) .
\end{align}
The terms symmetric under $\{ m,x \} \leftrightarrow \{ k,y \}$
 vanish, so
\begin{align}
   \Omega^{[2-2]}_n &= -\frac{(\pi L)^3}{3} \int \!\! dxdy C_{ijk}C_{jlm}
  \non
  & \mbox{} \times dP_0^l(x) \wedge dP_0^i(y) \left
  (\Xd{m}(x)P_0^k(x)  \deldash (x-y)  + \Xd{k}(y)P_0^m(y)  \deldash
  (x-y) \right) \non
  & =  \frac{(\pi L)^3}{3} \int \!\! dxdy C_{ikl}C_{jml}  \non
  & \mbox{} \times dP_0^i(x) \wedge dP_0^j(y) \left
  (\Xd{k}(x)P_0^m(x)  \deldash (x-y)  + \Xd{m}(y)P_0^k(y)  \deldash
  (x-y) \right).
\end{align}
Thus
\begin{equation}
 n  =  \frac{(\pi L)^3}{3} C_{ikl}C_{jml} \left[ \Xd{k}(x)P_0^m(x) +
  \Xd{m}(y)P_0^k(y) \right] \deldash (x-y) .
\end{equation}



\begin{thebibliography}{99}



\bibitem{CDS97} Alain Connes, Michael R.~Douglas, Albert Schwarz,
``Noncommutative Geometry and Matrix Theory: Compactification on
Tori,'' JHEP {\bf 9802} (1998) 003, {\tt hep-th/9711162}.

\bibitem{CH98} Chong-Sun Chu, Pei-Ming Ho,
``Noncommutative Open String and D-brane,'' Nucl.~Phys. {\bf B550}
(1999) 151-168, {\tt hep-th/9812219}.

\bibitem{IIBnoncomm}
H.~Aoki, N.~Ishibashi, S.~Iso, H.~Kawai, Y.~Kitazawa and T.~Tada,
``Noncommutative Yang-Mills in IIB matrix model,''
Nucl.~Phys. \textbf{B565} (2000) 176-192, \texttt{hep-th/9908141}.

\bibitem{BFSS} T.~Banks, W.~Fischler, S.~H.~Shenker, L.~Susskind,
``M Theory As A Matrix Model: A Conjecture,''
Phys.~Rev. {\bf D55} (1997) 5112-5128, {\tt hep-th/9610043}.



\bibitem{BST} E.~Bergshoeff, E.~Sezgin, P.~K.~Townsend,
``Supermembranes and Eleven-Dimensional Supergravity,''
Phys.~Lett. {\bf B189} (1987) 75-78. \\
   E.~Bergshoeff, E.~Sezgin, P.~K.~Townsend,
``Properties of the Eleven-Dimensional Supermembrane Theory,''
Ann.~Phys. {\bf 185} (1988) 330-368.



\bibitem{openMem}  Ph.~Brax, J.~Mourad,
``Open supermembranes in eleven dimensions,''  Phys.~Lett. {\bf B408} 
(1997) 142-150, {\tt hep-th/9704165};
``Open Supermembranes Coupled to M-Theory Five-Branes,''
Phys.~Lett. {\bf B416} (1998) 295-302, {\tt hep-th/9707246}. \\
 K.~Ezawa, Y.~Matsuo, K.~Murakami,
``Matrix Regularization of an Open Supermembrane ---towards M-theory
five-branes via open supermembranes ---,'' Phys.~Rev. {\bf D57} (1998)
5118-5133, {\tt hep-th/9707200}. \\
 Bernard de Wit, Kasper Peeters, Jan Plefka,
``Open and Closed Supermembranes with Winding,''
Nucl.~Phys.~Proc.~Suppl. {\bf 68} (1998) 206-215, 
{\tt  hep-th/9710215}. \\
 Martin Cederwall,
``Boundaries of 11-Dimensional Membranes,''
Mod.~Phys.~Lett. \textbf{A12} (1997) 2641-2645,
\texttt{hep-th/9704161}.


\bibitem{CH99} Chong-Sun Chu, Pei-Ming Ho,
``Constrained Quantization of Open String in Background B Field and
Noncommutative D-brane,'' to appear in Nucl.~Phys. {\bf B}, 
{\tt  hep-th/9906192}.

\bibitem{AAS-J99}  F.~Ardalan, H.~Arfaei, M.~M.~Sheikh-Jabbari,
``Dirac Quantization of Open Strings and Noncommutativity in
Branes,'' {\tt hep-th/9906161}.

\bibitem{BCasDC} M.~M.~Sheikh-Jabbari, A.~Shirzad,
``Boundary Conditions as Dirac Constraints,'' {\tt hep-th/9907055}.


\bibitem{KT00}
T.~Kawano and T.~Takahashi,
``Open string field theory on noncommutative space,''
\texttt{hep-th/9912274}.


\bibitem{Diractext} P.~A.~M.~Dirac,
``{\it Lectures on Quantum Mechanics\/}'' (Belfer Graduate School of
Science, Yeshiva Univ., 1964)

\bibitem{HRTtext} A.~Hanson, T.~Regge, C.~Teitelboim,
``Constrained Hamiltonian Systems'' (Accademia Nazionale dei Lincei 1976)

\bibitem{SW99}  Nathan Seiberg, Edward Witten,
``String Theory and Noncommutative Geometry,'' JHEP {\bf 9909} (1999)
032, {\tt hep-th/9908142}.


\bibitem{Strominger:1996ac}
A.~Strominger,
``Open p-branes,''
Phys.\ Lett.\  {\bf B383}, 44 (1996) ,
\texttt{hep-th/9512059}.


\bibitem{CS97} C.~S.~Chu, E.~Sezgin,
``M-Fivebrane from the Open Supermembrane,''
JHEP \textbf{9712} (1997) 001, \texttt{hep-th/9710223}.

\bibitem{BBSS00}  E.~Bergshoeff, D.~S.~Berman, J.~P.~van~der~Schaar,
  P.~Sundell,
``A Noncommutative M-Theory Five-brane,'' \texttt{hep-th/0005026}.


\end{thebibliography}
\end{document}